\documentclass[twocolumn]{aastex631}

\usepackage{threeparttable}
\usepackage{booktabs}
\usepackage{amsmath}

\begin{document}

\title{Merger Shocks Enhance Quenching in Local Galaxy Clusters}

\correspondingauthor{Ian D. Roberts}
\email{ianr@uwaterloo.ca}

\author[0000-0002-0692-0911]{Ian D. Roberts}
\affiliation{Department of Physics \& Astronomy, University of Waterloo, Waterloo, ON N2L 3G1, Canada}
\affiliation{Waterloo Centre for Astrophysics, University of Waterloo, 200 University Ave W, Waterloo, ON N2L 3G1, Canada}
\affiliation{Leiden Observatory, Leiden University, PO Box 9513, 2300 RA Leiden, The Netherlands}



\begin{abstract}
We report evidence for enhanced quenching in low-redshift galaxy clusters hosting radio relics. This effect is strongest for low-mass galaxies and is consistent with a rapid quenching of star formation. These results imply that merger shocks in the intracluster medium play a role in driving environmental quenching, which we argue is due to the elevated ram pressure experienced by satellite galaxies in these disturbed systems.
\end{abstract}



\section{Introduction} \label{sec:intro}

Galaxy clusters are the most-massive gravitationally bound objects in the Universe. The primary channel for the mass growth of clusters is hierarchical merging, both via accretion of lower mass groups, as well as major mergers with similar-mass clusters. Major cluster mergers are among the most energetic events in the Universe, capable of releasing energies of $10^{64}\,\mathrm{erg}$ or more \citep[e.g.][]{sarazin2002}. Mergers deposit new galaxies as well as drive shocks, increased turbulence, and sloshing motions through the intracluster medium (ICM). These effects can be observed through galaxy substructure and non-Gaussian velocity profiles \citep[e.g.][]{yahil1977,dressler1988,hou2009,martinez2012}, asymmetric X-ray distributions \citep[e.g.][]{rasia2013,weissmann2013}, and diffuse radio continuum emission in the form of radio relics and halos \citep[e.g.][]{feretti2012, vanweeren2019}.
\par
Given the increased velocity dispersions and ICM motions in disturbed systems, cluster mergers may be expected to impact the evolution of member galaxies. In particular, the dependence of ram pressure stripping on the relative velocity between galaxies and the ICM \cite[e.g.][]{gunn1972} implies that, all else equal, dynamical disturbances will increase the strength of ram pressure. Results on the effect of cluster dynamical state on galaxy evolution are mixed, with some studies finding an excess of star-forming galaxies in merging/disturbed clusters \citep[e.g.][]{ribeiro2010,carollo2013,stroe2015,roberts2017,stroe2017} and others finding a deficit \citep[e.g.][]{pranger2014,deshev2017}. Many (though not all) of these previous studies are based on classifications according the the group/cluster velocity distribution (i.e.\ Gaussian or non-Gaussian), however such classifications can be highly uncertain due to projection effects \citep{roberts2019b}.
\par
A far more reliable tracer of recent merger history is the presence of a radio relic\footnote{For the remainder of the paper, when referring to radio relics we are specifically referring to the \textit{radio gischt} class that is associated with merger shocks.}. Radio relics are diffuse synchrotron-emitting structures observed in the radio continuum. It is broadly accepted that these relics trace shock fronts from cluster mergers, in which cosmic ray electrons are accelerated by diffusive shock acceleration \citep[e.g.][]{blandford1978,jones1991,malkov2001}. Therefore, observing relics in clusters provides an avenue to generate samples of merging clusters with high purity.
\par
In this paper we compile all clusters in the Sloan Digital Sky Survey (SDSS) at low redshift with known radio relics. The purpose of this work is to address one simple, but important, question: does the presence of a merger shock impact star formation quenching in low-redshift clusters? As outlined above, constraints on this question will provide insight into the physical mechanisms driving environmental quenching in clusters.
\par
The outline of this paper is as follows. In Sect.~\ref{sec:data} we describe our cluster and galaxy samples, in Sect.~\ref{sec:results} we present observed quenched fractions for galaxies in clusters hosting a radio relic (relative to a control sample), and in Sect.~\ref{sec:discussion} we provide a discussion and conclusions. Throughout we assume a flat $\mathrm{\Lambda CDM}$ cosmology with $H_0 = 70\,\mathrm{km\,s^{-1}\,Mpc^{-1}}$, $\Omega_M = 0.3$, and $\Omega_\Lambda = 0.7$.

\section{Data \& Methods} \label{sec:data}

\subsection{Cluster Sample} \label{sec:clusters}

\begin{table*}[!ht]
\begin{threeparttable}
\centering
\caption{Clusters hosting a radio relic}
\label{tab:relic_table}
\begin{tabular}{l c c c c c c}
\toprule
\toprule
Name & Y07 ID\tnote{1} & Redshift\tnote{2} & R.A.\tnote{2} & Decl.\tnote{2} & $\log\,(M_h/\mathrm{M_\odot})$\tnote{3} & $N_\mathrm{gal}$\tnote{4} \\
\midrule
Abell 1367           & 3 & 0.0225  & 11h44m44.60s    & +19d41m59.0s   & 14.6 & 168 \\
Abell 1656           & 1 & 0.0234  & 12h59m44.40s    & +27d54m44.9s   & 14.9 & 700 \\
Abell 0168           & 25 & 0.0451  & 01h15m07.58s    & +00d19m10.8s   & 14.5 & 172 \\
Abell 1904           & 26 & 0.0718  & 14h22m13.21s    & +48d35m59.7s   & 14.8 & 156 \\ 
RXCJ1053.7+5452      & 146 & 0.0720  & 10h53m50.83s    & +54d49m55.1s   & 14.4 & 55 \\
Abell 2061           & 29 & 0.0783  & 15h21m19.90s    & +30d37m13.0s   & 14.9 & 166 \\        
Abell 2255           & 32 & 0.0801  & 17h12m50.04s    & +64d03m10.6s   & 15.0 & 231 \\
Abell 2249           & 218 & 0.0849  & 17h09m36.76s    & +34d28m04.1s   & 14.5 & 109 \\
Abell 0610           & 277 & 0.0970  & 07h59m21.77s    & +27d08m29.2s   & 14.4 & 44 \\             
\bottomrule
\end{tabular}
\begin{tablenotes}
    \item [1] modelC\_group catalogue from \citet{yang2007}
    \item [2] \href{https://simbad.unistra.fr/simbad/}{SIMBAD database}
    \item [3] Abundance matching from \citet{yang2007}
    \item [4] Spectroscopic member galaxies according to criteria in Sect.~\ref{sec:galaxies}
\end{tablenotes}
\end{threeparttable}
\end{table*}

We select all galaxy clusters at $z < 0.1$ with known radio relics. Clusters hosting radio relics are drawn from \citet{vanweeren2019} and \citet{botteon2022}. From these compilations we select all clusters at $z < 0.1$ that are classified as hosting either a confirmed or a candidate radio relic. Only one cluster in our sample, A2255, is classified as a candidate radio relic \citep{vanweeren2019}. We also require that the clusters are within the footprint of the main SDSS spectroscopic sample. This gives 9 clusters that form our radio relic sample listed in Table~\ref{tab:relic_table}.
\par
For comparison, we also define a control sample of galaxy clusters drawn from the \citet{yang2007} SDSS group catalogue. We restrict to clusters that fall within the redshift and halo mass range of our radio relic sample ($z < 0.1$ and $14.4 \le \log\,[M_h\,/\,\mathrm{M_\odot}] \le 15.0$). After removing the systems that make up our radio relic sample, this leaves 121 control clusters. Given the relative rarity of relics in clusters \citep[e.g.][]{botteon2022}, it is almost certain that this control sample is dominated by clusters that do not host a radio relic, though this has not been explicitly confirmed (and the requisite data to do so does not exist for all clusters in the control sample).

\subsection{Galaxy Sample} \label{sec:galaxies}

\begin{figure}[!ht]
    \centering
    \includegraphics[width = \columnwidth]{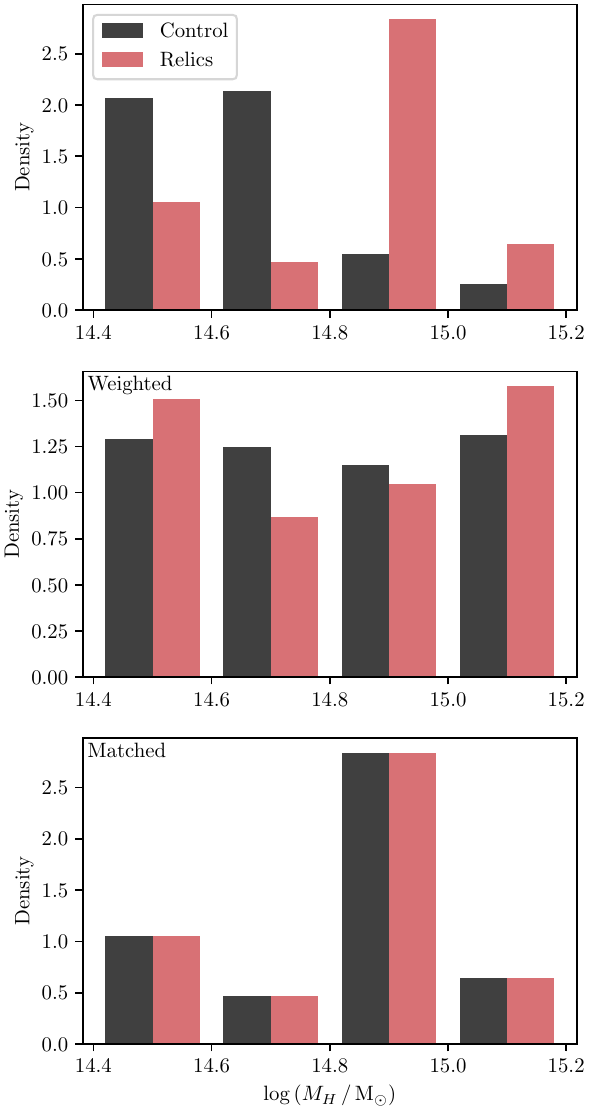}
    \caption{Distributions of host halo mass for galaxies in the radio relic (red) and control (black) samples. Top panel shows the unweighted distributions. Middle panel shows distributions where galaxies are weighted in order to match the halo mass distributions for the two samples (see Eq.~\ref{eq:weighting}). Bottom panel shows distributions for the radio relic sample and a subset of the control sample that is matched according to halo mass.}
    \label{fig:Mh_dist}
\end{figure}

We select member galaxies for each of our clusters based on offsets in projected radius and velocity from the cluster centre. We consider galaxies to be members of a given cluster if they are within $R_{180}$ and have a velocity offset of less than $3000\,\mathrm{km\,s^{-1}}$. The qualitative results are unchanged if instead we use a velocity offset of $<3\sigma_v$. We follow \citet{yang2007} and calculate $R_{180}$ as
\begin{equation}
    R_{180} = 1.26\,h^{-1}\,\mathrm{Mpc} \left(\frac{M_h}{10^{14}\,h^{-1}\,\mathrm{M_\odot}}\right)^{1/3}\,(1 + z_\mathrm{clust})^{-1},
\end{equation}
\noindent
where $z_\mathrm{clust}$ is the cluster redshift. The parent catalogue from which galaxies are selected is the GALEX-SDSS-WISE Legacy Catalog (GSWLC, \citealt{salim2016,salim2018}). We also take stellar masses and star formation rate estimates for each member galaxy from the GSWLC and we only include galaxies with $\log\,(M_\star\,/\,\mathrm{M_\odot}) \ge 9.5$. In total, this matching results in 1801 galaxies in the radio relic sample and 13215 galaxies in the control sample.
\par
The median galaxy redshifts for the radio relic and control galaxy samples are 0.05 and 0.07, respectively, which correspond to an SDSS stellar mass completeness of $\log\,(M_\star\,/\,\mathrm{M_\odot}) \simeq 10$ \citep{weigel2016}. Thus we are not stellar mass complete over our full mass range, however both the radio relic and control samples have similar redshift distributions and thus should be similarly incomplete. We note that we also compute quenched fractions in narrow bins of stellar mass in order to further mitigate any stellar mass dependencies.
\par
While the radio relic and control clusters span the same range in halo mass, the relic-hosting clusters are skewed towards larger halo mass than the control (see Fig.~\ref{fig:Mh_dist}, top). As quenched fractions correlate with host halo mass \citep[e.g.][]{wetzel2012}, we take two different approaches to ensuring that this halo-mass bias is not driving the results presented in this work.
\par
Our first approach is to re-weight the galaxies in each sample in order to ensure a fair comparison. Weights are derived in narrow bins of halo mass:
\begin{multline}
    \log\,(M_h\,/\,\mathrm{M_\odot}) = [14.4,\,14.6),\;[14.6,\,14.8),\\\;[14.8,\,15.0),\;[15.0,\,15.2).
\end{multline}
\noindent
For each halo mass bin, $i$, the weight is calculated as
\begin{equation} \label{eq:weighting}
    w_i = \frac{N_\mathrm{total}}{N_i},
\end{equation}
\noindent
where $N_\mathrm{total}$ is the total number of galaxies for a given sample and $N_i$ is the number of galaxies for that sample which fall within the halo mass bin, $i$. Weights are calculated separately for the control and radio relic galaxy samples, and when applied have the effect of enforcing a roughly uniform halo mass distribution over our sample range. We also experimented with different weighting schemes, as well as applying no weighting at all. In no cases did these variations affect the scientific conclusions from this work.
\par
Our second approach is to construct a smaller control sample that is explicitly matched to the halo mass distribution for the radio relic sample. In the same halo mass bins as used for the weighting, we randomly draw a galaxy from the control sample to be matched with each galaxy from the relic sample in the same halo mass bin. This gives a control sample that is of equal size and matched in halo mass to the relic sample. When doing computations with the matched control sample, we always take the median of 1000 random realizations of the matched control sample in order to account for the random nature of the matching procedure.
\par
In Fig.~\ref{fig:Mh_dist} (middle and bottom) we show the halo mass distributions resulting from the sample weighting and the sample matching described above. In both cases, the result is a control sample that is well matched to the halo mass distribution of the radio relic sample.

\subsection{Quenched Fractions} \label{sec:qfrac}

\begin{figure*}[!ht]
    \centering
    \includegraphics[width = 0.48\textwidth]{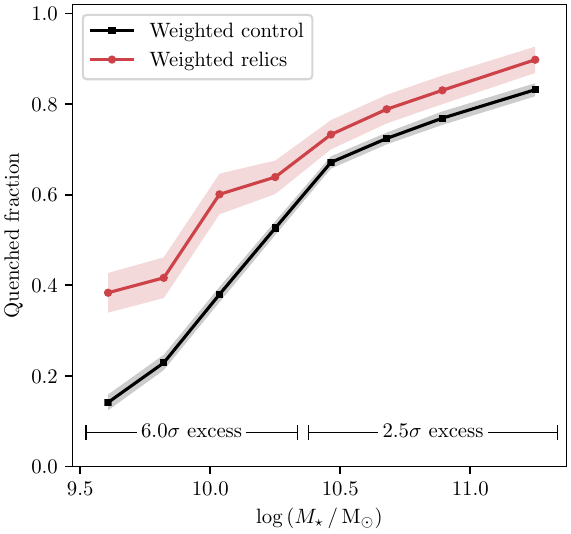}
    \hfill
    \includegraphics[width = 0.48\textwidth]{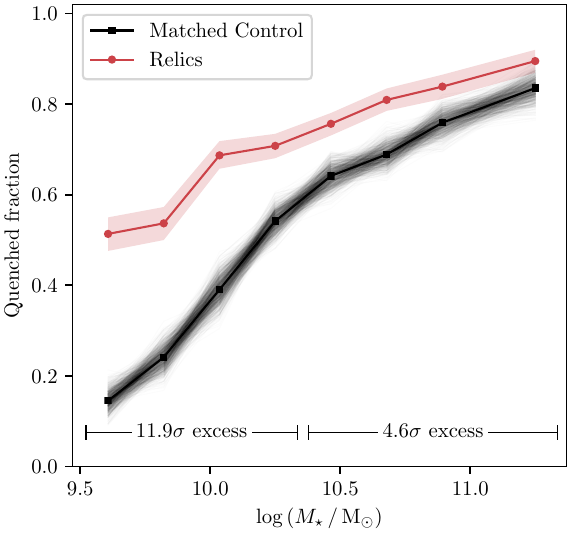}
    \caption{\textit{Left:} Weighted quenched fraction as a function of stellar mass for galaxies in the radio relic (red) and control (black) samples. The vertical shading corresponds to the 16 and 84 percentile confidence interval on the quenched fraction, derived via bootstrap re-sampling. \textit{Right:} Quenched fraction as a function of stellar mass for galaxies in the radio relic (red) and matched control (black) samples. For the relic sample, the vertical shading again corresponds to the 16 and 84 percentile bootstrap confidence interval on the quenched fraction. For the matched control sample, the faded lines correspond to the individual trends for each random realization of the matching procedure (see text for details), and the heavier line-type marks the median of these realizations.}
    \label{fig:qfrac}
\end{figure*}

In general, the quenched fraction for a given set of galaxies is calculated as
\begin{equation} \label{eq:qfrac}
    f_Q = \frac{N_\mathrm{Q}}{N},
\end{equation}
\noindent
where $N_\mathrm{Q}$ and $N$ are the number of quenched galaxies and the total number of galaxies, respectively. To identify quenched galaxies we follow numerous previous works \citep[e.g.][]{wetzel2012} and use a single cut in specific star formation rate ($\mathrm{sSFR} = \mathrm{SFR} / M_\star$) at $\log (\mathrm{sSFR / yr^{-1}}) = -11$.
\par
We calculate quenched fractions in narrow bins of stellar mass. When comparing the radio relic and matched control samples we compute quenched fractions according to Eq.~\ref{eq:qfrac}. We also separately apply the weights described in Sect.~\ref{sec:galaxies}, and in this case the weighted quenched fraction is given by
\begin{equation} \label{eq:qfrac_w}
    f_Q = \frac{\sum w\,(\mathrm{\log sSFR} < -11)}{\sum w}.
\end{equation}

\section{Results} \label{sec:results}

In Fig.~\ref{fig:qfrac} we show the main scientific result of this paper, galaxies within clusters with radio relics have enhanced quenched fractions relative to the control cluster sample. This enhancement is most clear at the low-mass end, consistent with many previous works showing that environmental quenching effects preferentially impact low-mass galaxies (at least at low-$z$, e.g.\ \citealt{haines2006,bamford2009,peng2010,roberts2019}). In the lowest-mass bins the quenched fraction in the radio relic sample is at least a factor of two above the control. These statements hold regardless of whether we consider the weighted samples (left) or the matched control sample (right).
\par
For low-mass galaxies ($9.5 \le \log[M_\star / \mathrm{M_\odot}] < 10.5$), the quenched fraction shows a $6\sigma$ and $12\sigma$ excess relative to the control sample, for the weighted and matched samples respectively. For the matched approach, we measure the excess relative to the median of the random matches. Conservatively, we conclude that galaxies in the radio relic sample have enhanced quenched fractions at greater than $6\sigma$ significance. For high-mass galaxies ($10.5 \le \log[M_\star / \mathrm{M_\odot}] < 11.5$), this excess is weaker in magnitude and at lesser significance ($2-4\sigma$).

\begin{figure}
    \centering
    \includegraphics[width=\columnwidth]{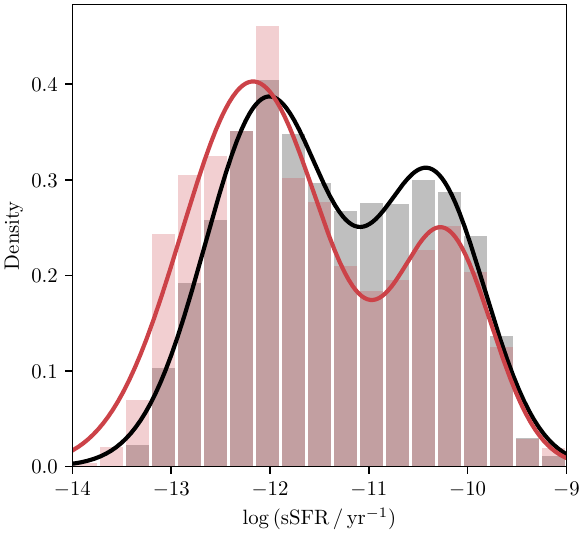}
    \caption{Specific star formation rate distributions for the control and radio relic samples. Both distributions have been weighted according to Eq.~\ref{eq:weighting}. Observed distributions are shown by histograms and we also overlay the best fit double-Gaussian distribution for each each sample.}
    \label{fig:ssfr}
\end{figure}

In Fig.~\ref{fig:ssfr} we show the sSFR distributions for the relic and control samples, including a double-Gaussian fit for each sample. The distributions in Fig.~\ref{fig:ssfr} are weighted according to Eq.~\ref{eq:weighting}, though we note that the same conclusions hold when instead considering the matched control and unweighted relic samples. This again shows the excess of quenched galaxies in the radio relic clusters. The shape of the high-sSFR Gaussian component, corresponding to the star-forming population, is nearly identical for the control and radio relic samples (modulo normalization). This is consistent with the difference between the two samples being driven by a rapid quenching process where the quenching timescale is short relative to the gas depletion timescale (see e.g. \citealt{wetzel2012}).

\section{Discussion \& Conclusions} \label{sec:discussion}

The excess in quenched galaxies for clusters with radio relics has important implications for the physical drivers of environmental quenching. The efficiency with which galaxies are quenched appears to be enhanced in the presence of merger-induced motions and shock waves in the ICM. The most commonly invoked mechanisms for explaining environmental quenching in low-$z$ clusters are the ram-pressure stripping of cold gas and starvation (see \citealt{cortese2021,boselli2022_review} for recent reviews). Starvation is sometimes sub-divided into gas consumption associated with the stripping of the circumgalactic medium (CGM), versus gas consumption in association with the inability of new gas to cool and condense out of the CGM due to the high virial temperature of the cluster. The former is simply a manifestation of a moderate ram pressure whereas the latter is more physically distinct.
\par
Because ram pressure has a quadratic dependence on the relative velocity between the galaxy and the ICM, the strength of ram pressure should be enhanced (all else equal) in the presence of sloshing motions, and especially in the vicinity of shocks moving through the ICM. Thus the excess quenched fraction shown in Fig.~\ref{fig:qfrac} favours a ram pressure like quenching mechanism, with the cold ISM and/or the CGM being removed, as opposed to a starvation scenario tied to the lack of gas cooling.
\par
This is not the first time that quenching, particularly via ram pressure stripping, has been associated with diffuse radio relics in clusters. The best known example would be Abell 1367 (a member of our radio relic cluster sample), where three prominent jellyfish galaxies are spatially coincident with the radio relic to the northwest of the cluster center \citep{gavazzi1987_a1367,gavazzi2001,roberts2021_LOFARclust}. It is both thought that the merger shock has contributed to the extreme stripped tails observed for these galaxies, and that the radio continuum tails in turn feed the radio relic with seed cosmic ray electrons \citep{chong2019}. Unfortunately, the number of known jellyfish galaxies in clusters hosting radio relics is small (likely ${\lesssim}50$), leaving any attempted association between relics and stripped tails with extremely limited statistical power. 
\par
At intermediate redshift, \citet{stroe2015,stroe2017} show that merging clusters (as traced by radio haloes and/or relics) host an excess of $\mathrm{H\alpha}$ emitters relative to relaxed clusters (and the field). This is attributed to gas compression from ICM shocks catalyzing star formation and is consistent with works at low redshift finding enhanced SFRs in galaxies experiencing RPS \citep[e.g.][]{gavazzi2001,vulcani2018_sf,roberts2020,roberts2022_lofar_manga}. At similar redshifts, \citet{ebeling2014,ebeling2019} and \citet{mcpartland2016} find a number of jellyfish galaxies with tails of extra-planar material which they argue are preferentially associated with ongoing cluster mergers. This rapid gas consumption, in concert with gas removal via increased ram pressure, may lead to rapid quenching and thus the excess of quenched galaxies that we observe at $z \sim 0$.
\par
This work clearly shows excess environmental quenching associated with the presence of ICM shocks. With more radio relic detections moving forward, for example with the completion of the LOFAR Two-metre Sky Survey \citep{shimwell2017,shimwell2019,shimwell2022} as well as future surveys from the Square Kilometre Array, these conclusions can be strengthened further. An ultimate goal will be to directly test whether a quantitative association is present between ram pressure tailed galaxies and the location (and path) of merger shocks in clusters.

\begin{acknowledgments}
We thank Laura Parker and Mike Hudson for giving feedback on an early draft. IDR acknowledges support from the Banting Fellowship Program.
\end{acknowledgments}

%

\vspace{5mm}
\facilities{Apache Point Observatory}


\software{AstroPy \citep{astropy2013}, Matplotlib \citep{hunter2007}, NumPy \citep{harris2020}}




\bibliography{main}{}

\begin{thebibliography}{}
\expandafter\ifx\csname natexlab\endcsname\relax\def\natexlab#1{#1}\fi
\providecommand{\url}[1]{\href{#1}{#1}}
\providecommand{\dodoi}[1]{doi:~\href{http://doi.org/#1}{\nolinkurl{#1}}}
\providecommand{\doeprint}[1]{\href{http://ascl.net/#1}{\nolinkurl{http://ascl.net/#1}}}
\providecommand{\doarXiv}[1]{\href{https://arxiv.org/abs/#1}{\nolinkurl{https://arxiv.org/abs/#1}}}

\bibitem[{{Astropy Collaboration} {et~al.}(2013){Astropy Collaboration}, {Robitaille}, {Tollerud}, {Greenfield}, {Droettboom}, {Bray}, {Aldcroft}, {Davis}, {Ginsburg}, {Price-Whelan}, {Kerzendorf}, {Conley}, {Crighton}, {Barbary}, {Muna}, {Ferguson}, {Grollier}, {Parikh}, {Nair}, {Unther}, {Deil}, {Woillez}, {Conseil}, {Kramer}, {Turner}, {Singer}, {Fox}, {Weaver}, {Zabalza}, {Edwards}, {Azalee Bostroem}, {Burke}, {Casey}, {Crawford}, {Dencheva}, {Ely}, {Jenness}, {Labrie}, {Lim}, {Pierfederici}, {Pontzen}, {Ptak}, {Refsdal}, {Servillat}, \& {Streicher}}]{astropy2013}
{Astropy Collaboration}, {Robitaille}, T.~P., {Tollerud}, E.~J., {et~al.} 2013, A\&A, 558, A33, \dodoi{10.1051/0004-6361/201322068}

\bibitem[{{Bamford} {et~al.}(2009){Bamford}, {Nichol}, {Baldry}, {Land}, {Lintott}, {Schawinski}, {Slosar}, {Szalay}, {Thomas}, {Torki}, {Andreescu}, {Edmondson}, {Miller}, {Murray}, {Raddick}, \& {Vandenberg}}]{bamford2009}
{Bamford}, S.~P., {Nichol}, R.~C., {Baldry}, I.~K., {et~al.} 2009, MNRAS, 393, 1324, \dodoi{10.1111/j.1365-2966.2008.14252.x}

\bibitem[{{Blandford} \& {Ostriker}(1978)}]{blandford1978}
{Blandford}, R.~D., \& {Ostriker}, J.~P. 1978, \apjl, 221, L29, \dodoi{10.1086/182658}

\bibitem[{{Boselli} {et~al.}(2022){Boselli}, {Fossati}, \& {Sun}}]{boselli2022_review}
{Boselli}, A., {Fossati}, M., \& {Sun}, M. 2022, A\&AR, 30, 3, \dodoi{10.1007/s00159-022-00140-3}

\bibitem[{{Botteon} {et~al.}(2022){Botteon}, {Shimwell}, {Cassano}, {Cuciti}, {Zhang}, {Bruno}, {Camillini}, {Natale}, {Jones}, {Gastaldello}, {Simionescu}, {Rossetti}, {Akamatsu}, {van Weeren}, {Brunetti}, {Br{\"u}ggen}, {Groeneveld}, {Hoang}, {Hardcastle}, {Ignesti}, {Di Gennaro}, {Bonafede}, {Drabent}, {R{\"o}ttgering}, {Hoeft}, \& {de Gasperin}}]{botteon2022}
{Botteon}, A., {Shimwell}, T.~W., {Cassano}, R., {et~al.} 2022, \aap, 660, A78, \dodoi{10.1051/0004-6361/202143020}

\bibitem[{{Carollo} {et~al.}(2013){Carollo}, {Cibinel}, {Lilly}, {Miniati}, {Norberg}, {Silverman}, {van Gorkom}, {Cameron}, {Finoguenov}, {Peng}, {Pipino}, \& {Rudick}}]{carollo2013}
{Carollo}, C.~M., {Cibinel}, A., {Lilly}, S.~J., {et~al.} 2013, ApJ, 776, 71, \dodoi{10.1088/0004-637X/776/2/71}

\bibitem[{{Cortese} {et~al.}(2021){Cortese}, {Catinella}, \& {Smith}}]{cortese2021}
{Cortese}, L., {Catinella}, B., \& {Smith}, R. 2021, PASA, 38, e035, \dodoi{10.1017/pasa.2021.18}

\bibitem[{{Deshev} {et~al.}(2017){Deshev}, {Finoguenov}, {Verdugo}, {Ziegler}, {Park}, {Hwang}, {Haines}, {Kamphuis}, {Tamm}, {Einasto}, {Hwang}, \& {Park}}]{deshev2017}
{Deshev}, B., {Finoguenov}, A., {Verdugo}, M., {et~al.} 2017, \aap, 607, A131, \dodoi{10.1051/0004-6361/201731235}

\bibitem[{{Dressler} \& {Shectman}(1988)}]{dressler1988}
{Dressler}, A., \& {Shectman}, S.~A. 1988, AJ, 95, 985, \dodoi{10.1086/114694}

\bibitem[{{Ebeling} \& {Kalita}(2019)}]{ebeling2019}
{Ebeling}, H., \& {Kalita}, B.~S. 2019, \apj, 882, 127, \dodoi{10.3847/1538-4357/ab35d6}

\bibitem[{{Ebeling} {et~al.}(2014){Ebeling}, {Stephenson}, \& {Edge}}]{ebeling2014}
{Ebeling}, H., {Stephenson}, L.~N., \& {Edge}, A.~C. 2014, ApJL, 781, L40, \dodoi{10.1088/2041-8205/781/2/L40}

\bibitem[{{Feretti} {et~al.}(2012){Feretti}, {Giovannini}, {Govoni}, \& {Murgia}}]{feretti2012}
{Feretti}, L., {Giovannini}, G., {Govoni}, F., \& {Murgia}, M. 2012, \aapr, 20, 54, \dodoi{10.1007/s00159-012-0054-z}

\bibitem[{{Gavazzi} {et~al.}(2001){Gavazzi}, {Boselli}, {Mayer}, {Iglesias-Paramo}, {V{\'\i}lchez}, \& {Carrasco}}]{gavazzi2001}
{Gavazzi}, G., {Boselli}, A., {Mayer}, L., {et~al.} 2001, ApJL, 563, L23, \dodoi{10.1086/338389}

\bibitem[{{Gavazzi} \& {Jaffe}(1987)}]{gavazzi1987_a1367}
{Gavazzi}, G., \& {Jaffe}, W. 1987, A\&A, 186, L1

\bibitem[{{Ge} {et~al.}(2019){Ge}, {Sun}, {Liu}, {Rudnick}, {Sarazin}, {Forman}, {Jones}, {Chen}, {Liu}, {Yagi}, {Boselli}, {Fossati}, \& {Gavazzi}}]{chong2019}
{Ge}, C., {Sun}, M., {Liu}, R.-Y., {et~al.} 2019, \mnras, 486, L36, \dodoi{10.1093/mnrasl/slz049}

\bibitem[{{Gunn} \& {Gott}(1972)}]{gunn1972}
{Gunn}, J.~E., \& {Gott}, III, J.~R. 1972, ApJ, 176, 1, \dodoi{10.1086/151605}

\bibitem[{{Haines} {et~al.}(2006){Haines}, {La Barbera}, {Mercurio}, {Merluzzi}, \& {Busarello}}]{haines2006}
{Haines}, C.~P., {La Barbera}, F., {Mercurio}, A., {Merluzzi}, P., \& {Busarello}, G. 2006, ApJL, 647, L21, \dodoi{10.1086/507297}

\bibitem[{Harris {et~al.}(2020)Harris, Millman, van~der Walt, Gommers, Virtanen, Cournapeau, Wieser, Taylor, Berg, Smith, Kern, Picus, Hoyer, van Kerkwijk, Brett, Haldane, del R{\'{i}}o, Wiebe, Peterson, G{\'{e}}rard-Marchant, Sheppard, Reddy, Weckesser, Abbasi, Gohlke, \& Oliphant}]{harris2020}
Harris, C.~R., Millman, K.~J., van~der Walt, S.~J., {et~al.} 2020, Nature, 585, 357, \dodoi{10.1038/s41586-020-2649-2}

\bibitem[{{Hou} {et~al.}(2009){Hou}, {Parker}, {Harris}, \& {Wilman}}]{hou2009}
{Hou}, A., {Parker}, L.~C., {Harris}, W.~E., \& {Wilman}, D.~J. 2009, ApJ, 702, 1199, \dodoi{10.1088/0004-637X/702/2/1199}

\bibitem[{Hunter(2007)}]{hunter2007}
Hunter, J.~D. 2007, Computing In Science \& Engineering, 9, 90, \dodoi{10.1109/MCSE.2007.55}

\bibitem[{{Jones} \& {Ellison}(1991)}]{jones1991}
{Jones}, F.~C., \& {Ellison}, D.~C. 1991, \ssr, 58, 259, \dodoi{10.1007/BF01206003}

\bibitem[{{Malkov} \& {Drury}(2001)}]{malkov2001}
{Malkov}, M.~A., \& {Drury}, L.~O. 2001, Reports on Progress in Physics, 64, 429, \dodoi{10.1088/0034-4885/64/4/201}

\bibitem[{{Mart{\'{i}}nez} \& {Zandivarez}(2012)}]{martinez2012}
{Mart{\'{i}}nez}, H.~J., \& {Zandivarez}, A. 2012, MNRAS, 419, L24, \dodoi{10.1111/j.1745-3933.2011.01170.x}

\bibitem[{{McPartland} {et~al.}(2016){McPartland}, {Ebeling}, {Roediger}, \& {Blumenthal}}]{mcpartland2016}
{McPartland}, C., {Ebeling}, H., {Roediger}, E., \& {Blumenthal}, K. 2016, MNRAS, 455, 2994, \dodoi{10.1093/mnras/stv2508}

\bibitem[{{Peng} {et~al.}(2010){Peng}, {Lilly}, {Kova{\v c}}, {Bolzonella}, {Pozzetti}, {Renzini}, {Zamorani}, {Ilbert}, {Knobel}, {Iovino}, {Maier}, {Cucciati}, {Tasca}, {Carollo}, {Silverman}, {Kampczyk}, {de Ravel}, {Sanders}, {Scoville}, {Contini}, {Mainieri}, {Scodeggio}, {Kneib}, {Le F{\`e}vre}, {Bardelli}, {Bongiorno}, {Caputi}, {Coppa}, {de la Torre}, {Franzetti}, {Garilli}, {Lamareille}, {Le Borgne}, {Le Brun}, {Mignoli}, {Perez Montero}, {Pello}, {Ricciardelli}, {Tanaka}, {Tresse}, {Vergani}, {Welikala}, {Zucca}, {Oesch}, {Abbas}, {Barnes}, {Bordoloi}, {Bottini}, {Cappi}, {Cassata}, {Cimatti}, {Fumana}, {Hasinger}, {Koekemoer}, {Leauthaud}, {Maccagni}, {Marinoni}, {McCracken}, {Memeo}, {Meneux}, {Nair}, {Porciani}, {Presotto}, \& {Scaramella}}]{peng2010}
{Peng}, Y.-j., {Lilly}, S.~J., {Kova{\v c}}, K., {et~al.} 2010, ApJ, 721, 193, \dodoi{10.1088/0004-637X/721/1/193}

\bibitem[{{Pranger} {et~al.}(2014){Pranger}, {B{\"o}hm}, {Ferrari}, {Maurogordato}, {Benoist}, {H{\"o}ller}, \& {Schindler}}]{pranger2014}
{Pranger}, F., {B{\"o}hm}, A., {Ferrari}, C., {et~al.} 2014, \aap, 570, A40, \dodoi{10.1051/0004-6361/201424727}

\bibitem[{{Rasia} {et~al.}(2013){Rasia}, {Meneghetti}, \& {Ettori}}]{rasia2013}
{Rasia}, E., {Meneghetti}, M., \& {Ettori}, S. 2013, The Astronomical Review, 8, 40, \dodoi{10.1080/21672857.2013.11519713}

\bibitem[{{Ribeiro} {et~al.}(2010){Ribeiro}, {Lopes}, \& {Trevisan}}]{ribeiro2010}
{Ribeiro}, A.~L.~B., {Lopes}, P.~A.~A., \& {Trevisan}, M. 2010, MNRAS, 409, L124, \dodoi{10.1111/j.1745-3933.2010.00962.x}

\bibitem[{{Roberts} \& {Parker}(2017)}]{roberts2017}
{Roberts}, I.~D., \& {Parker}, L.~C. 2017, MNRAS, 467, 3268, \dodoi{10.1093/mnras/stx317}

\bibitem[{{Roberts} \& {Parker}(2019)}]{roberts2019b}
---. 2019, MNRAS, 2291, \dodoi{10.1093/mnras/stz2666}

\bibitem[{{Roberts} \& {Parker}(2020)}]{roberts2020}
---. 2020, MNRAS, 495, 554, \dodoi{10.1093/mnras/staa1213}

\bibitem[{{Roberts} {et~al.}(2019){Roberts}, {Parker}, {Brown}, {Joshi}, {Hlavacek-Larrondo}, \& {Wadsley}}]{roberts2019}
{Roberts}, I.~D., {Parker}, L.~C., {Brown}, T., {et~al.} 2019, ApJ, 873, 42, \dodoi{10.3847/1538-4357/ab04f7}

\bibitem[{{Roberts} {et~al.}(2021){Roberts}, {van Weeren}, {McGee}, {Botteon}, {Drabent}, {Ignesti}, {Rottgering}, {Shimwell}, \& {Tasse}}]{roberts2021_LOFARclust}
{Roberts}, I.~D., {van Weeren}, R.~J., {McGee}, S.~L., {et~al.} 2021, \aap, 650, A111, \dodoi{10.1051/0004-6361/202140784}

\bibitem[{{Roberts} {et~al.}(2022){Roberts}, {Lang}, {Trotsenko}, {Bemis}, {Ellison}, {Lin}, {Pan}, {Ignesti}, {Leslie}, \& {van Weeren}}]{roberts2022_lofar_manga}
{Roberts}, I.~D., {Lang}, M., {Trotsenko}, D., {et~al.} 2022, ApJ, 941, 77, \dodoi{10.3847/1538-4357/ac9e9f}

\bibitem[{{Salim} {et~al.}(2018){Salim}, {Boquien}, \& {Lee}}]{salim2018}
{Salim}, S., {Boquien}, M., \& {Lee}, J.~C. 2018, ApJ, 859, 11, \dodoi{10.3847/1538-4357/aabf3c}

\bibitem[{{Salim} {et~al.}(2016){Salim}, {Lee}, {Janowiecki}, {da Cunha}, {Dickinson}, {Boquien}, {Burgarella}, {Salzer}, \& {Charlot}}]{salim2016}
{Salim}, S., {Lee}, J.~C., {Janowiecki}, S., {et~al.} 2016, ApJS, 227, 2, \dodoi{10.3847/0067-0049/227/1/2}

\bibitem[{{Sarazin}(2002)}]{sarazin2002}
{Sarazin}, C.~L. 2002, in Astrophysics and Space Science Library, Vol. 272, Merging Processes in Galaxy Clusters, ed. L.~{Feretti}, I.~M. {Gioia}, \& G.~{Giovannini}, 1--38, \dodoi{10.1007/0-306-48096-4_1}

\bibitem[{{Shimwell} {et~al.}(2017){Shimwell}, {R{\"o}ttgering}, {Best}, {Williams}, {Dijkema}, {de Gasperin}, {Hardcastle}, {Heald}, {Hoang}, {Horneffer}, {Intema}, {Mahony}, {Mandal}, {Mechev}, {Morabito}, {Oonk}, {Rafferty}, {Retana-Montenegro}, {Sabater}, {Tasse}, {van Weeren}, {Br{\"u}ggen}, {Brunetti}, {Chy{\.z}y}, {Conway}, {Haverkorn}, {Jackson}, {Jarvis}, {McKean}, {Miley}, {Morganti}, {White}, {Wise}, {van Bemmel}, {Beck}, {Brienza}, {Bonafede}, {Calistro Rivera}, {Cassano}, {Clarke}, {Cseh}, {Deller}, {Drabent}, {van Driel}, {Engels}, {Falcke}, {Ferrari}, {Fr{\"o}hlich}, {Garrett}, {Harwood}, {Heesen}, {Hoeft}, {Horellou}, {Israel}, {Kapi{\'n}ska}, {Kunert-Bajraszewska}, {McKay}, {Mohan}, {Orr{\'u}}, {Pizzo}, {Prandoni}, {Schwarz}, {Shulevski}, {Sipior}, {Smith}, {Sridhar}, {Steinmetz}, {Stroe}, {Varenius}, {van der Werf}, {Zensus}, \& {Zwart}}]{shimwell2017}
{Shimwell}, T.~W., {R{\"o}ttgering}, H.~J.~A., {Best}, P.~N., {et~al.} 2017, A\&A, 598, A104, \dodoi{10.1051/0004-6361/201629313}

\bibitem[{{Shimwell} {et~al.}(2019){Shimwell}, {Tasse}, {Hardcastle}, {Mechev}, {Williams}, {Best}, {R{\"o}ttgering}, {Callingham}, {Dijkema}, {de Gasperin}, {Hoang}, {Hugo}, {Mirmont}, {Oonk}, {Prandoni}, {Rafferty}, {Sabater}, {Smirnov}, {van Weeren}, {White}, {Atemkeng}, {Bester}, {Bonnassieux}, {Br{\"u}ggen}, {Brunetti}, {Chy{\.z}y}, {Cochrane}, {Conway}, {Croston}, {Danezi}, {Duncan}, {Haverkorn}, {Heald}, {Iacobelli}, {Intema}, {Jackson}, {Jamrozy}, {Jarvis}, {Lakhoo}, {Mevius}, {Miley}, {Morabito}, {Morganti}, {Nisbet}, {Orr{\'u}}, {Perkins}, {Pizzo}, {Schrijvers}, {Smith}, {Vermeulen}, {Wise}, {Alegre}, {Bacon}, {van Bemmel}, {Beswick}, {Bonafede}, {Botteon}, {Bourke}, {Brienza}, {Calistro Rivera}, {Cassano}, {Clarke}, {Conselice}, {Dettmar}, {Drabent}, {Dumba}, {Emig}, {En{\ss}lin}, {Ferrari}, {Garrett}, {G{\'e}nova-Santos}, {Goyal}, {G{\"u}rkan}, {Hale}, {Harwood}, {Heesen}, {Hoeft}, {Horellou}, {Jackson}, {Kokotanekov}, {Kondapally}, {Kunert-Bajraszewska}, {Mahatma}, {Mahony}, {Mandal}, {McKean},
  {Merloni}, {Mingo}, {Miskolczi}, {Mooney}, {Nikiel-Wroczy{\'n}ski}, {O'Sullivan}, {Quinn}, {Reich}, {Roskowi{\'n}ski}, {Rowlinson}, {Savini}, {Saxena}, {Schwarz}, {Shulevski}, {Sridhar}, {Stacey}, {Urquhart}, {van der Wiel}, {Varenius}, {Webster}, \& {Wilber}}]{shimwell2019}
{Shimwell}, T.~W., {Tasse}, C., {Hardcastle}, M.~J., {et~al.} 2019, A\&A, 622, A1, \dodoi{10.1051/0004-6361/201833559}

\bibitem[{{Shimwell} {et~al.}(2022){Shimwell}, {Hardcastle}, {Tasse}, {Best}, {R{\"o}ttgering}, {Williams}, {Botteon}, {Drabent}, {Mechev}, {Shulevski}, {van Weeren}, {Bester}, {Br{\"u}ggen}, {Brunetti}, {Callingham}, {Chy{\.z}y}, {Conway}, {Dijkema}, {Duncan}, {de Gasperin}, {Hale}, {Haverkorn}, {Hugo}, {Jackson}, {Mevius}, {Miley}, {Morabito}, {Morganti}, {Offringa}, {Oonk}, {Rafferty}, {Sabater}, {Smith}, {Schwarz}, {Smirnov}, {O'Sullivan}, {Vedantham}, {White}, {Albert}, {Alegre}, {Asabere}, {Bacon}, {Bonafede}, {Bonnassieux}, {Brienza}, {Bilicki}, {Bonato}, {Calistro Rivera}, {Cassano}, {Cochrane}, {Croston}, {Cuciti}, {Dallacasa}, {Danezi}, {Dettmar}, {Di Gennaro}, {Edler}, {En{\ss}lin}, {Emig}, {Franzen}, {Garc{\'\i}a-Vergara}, {Grange}, {G{\"u}rkan}, {Hajduk}, {Heald}, {Heesen}, {Hoang}, {Hoeft}, {Horellou}, {Iacobelli}, {Jamrozy}, {Jeli{\'c}}, {Kondapally}, {Kukreti}, {Kunert-Bajraszewska}, {Magliocchetti}, {Mahatma}, {Ma{\l}ek}, {Mandal}, {Massaro}, {Meyer-Zhao}, {Mingo}, {Mostert}, {Nair},
  {Nakoneczny}, {Nikiel-Wroczy{\'n}ski}, {Orr{\'u}}, {Pajdosz-{\'S}mierciak}, {Pasini}, {Prandoni}, {van Piggelen}, {Rajpurohit}, {Retana-Montenegro}, {Riseley}, {Rowlinson}, {Saxena}, {Schrijvers}, {Sweijen}, {Siewert}, {Timmerman}, {Vaccari}, {Vink}, {West}, {Wo{\l}owska}, {Zhang}, \& {Zheng}}]{shimwell2022}
{Shimwell}, T.~W., {Hardcastle}, M.~J., {Tasse}, C., {et~al.} 2022, A\&A, 659, A1, \dodoi{10.1051/0004-6361/202142484}

\bibitem[{{Stroe} {et~al.}(2017){Stroe}, {Sobral}, {Paulino-Afonso}, {Alegre}, {Calhau}, {Santos}, \& {van Weeren}}]{stroe2017}
{Stroe}, A., {Sobral}, D., {Paulino-Afonso}, A., {et~al.} 2017, \mnras, 465, 2916, \dodoi{10.1093/mnras/stw2939}

\bibitem[{{Stroe} {et~al.}(2015){Stroe}, {Sobral}, {Dawson}, {Jee}, {Hoekstra}, {Wittman}, {van Weeren}, {Br{\"u}ggen}, \& {R{\"o}ttgering}}]{stroe2015}
{Stroe}, A., {Sobral}, D., {Dawson}, W., {et~al.} 2015, \mnras, 450, 646, \dodoi{10.1093/mnras/stu2519}

\bibitem[{{van Weeren} {et~al.}(2019){van Weeren}, {de Gasperin}, {Akamatsu}, {Br{\"u}ggen}, {Feretti}, {Kang}, {Stroe}, \& {Zandanel}}]{vanweeren2019}
{van Weeren}, R.~J., {de Gasperin}, F., {Akamatsu}, H., {et~al.} 2019, \ssr, 215, 16, \dodoi{10.1007/s11214-019-0584-z}

\bibitem[{{Vulcani} {et~al.}(2018){Vulcani}, {Poggianti}, {Gullieuszik}, {Moretti}, {Tonnesen}, {Jaff{\'e}}, {Fritz}, {Fasano}, \& {Bettoni}}]{vulcani2018_sf}
{Vulcani}, B., {Poggianti}, B.~M., {Gullieuszik}, M., {et~al.} 2018, ApJL, 866, L25, \dodoi{10.3847/2041-8213/aae68b}

\bibitem[{{Weigel} {et~al.}(2016){Weigel}, {Schawinski}, \& {Bruderer}}]{weigel2016}
{Weigel}, A.~K., {Schawinski}, K., \& {Bruderer}, C. 2016, MNRAS, 459, 2150, \dodoi{10.1093/mnras/stw756}

\bibitem[{{Wei{\ss}mann} {et~al.}(2013){Wei{\ss}mann}, {B{\"o}hringer}, {{\v S}uhada}, \& {Ameglio}}]{weissmann2013}
{Wei{\ss}mann}, A., {B{\"o}hringer}, H., {{\v S}uhada}, R., \& {Ameglio}, S. 2013, A\&A, 549, A19, \dodoi{10.1051/0004-6361/201219333}

\bibitem[{{Wetzel} {et~al.}(2012){Wetzel}, {Tinker}, \& {Conroy}}]{wetzel2012}
{Wetzel}, A.~R., {Tinker}, J.~L., \& {Conroy}, C. 2012, MNRAS, 424, 232, \dodoi{10.1111/j.1365-2966.2012.21188.x}

\bibitem[{{Yahil} \& {Vidal}(1977)}]{yahil1977}
{Yahil}, A., \& {Vidal}, N.~V. 1977, ApJ, 214, 347, \dodoi{10.1086/155257}

\bibitem[{{Yang} {et~al.}(2007){Yang}, {Mo}, {van den Bosch}, {Pasquali}, {Li}, \& {Barden}}]{yang2007}
{Yang}, X., {Mo}, H.~J., {van den Bosch}, F.~C., {et~al.} 2007, ApJ, 671, 153, \dodoi{10.1086/522027}

\end{thebibliography}
\bibliographystyle{aasjournal}



\end{document}